\documentclass[aps,prb,reprint,groupedaddress]{revtex4-2}
\usepackage{amsmath}
\usepackage{bm} % bold math
\usepackage{amssymb}
\usepackage{color}
\usepackage{multirow}
\usepackage{url}
\usepackage[colorlinks,bookmarks=false,citecolor=red,linkcolor=blue,urlcolor=blue]{hyperref}
\usepackage{graphicx}
\usepackage{mathrsfs,bbm}
\usepackage{wrapfig}
\usepackage{physics}
\usepackage{enumerate}
\usepackage[version=3]{mhchem}%
\usepackage{gensymb}
\def\ahat{\hat{a}}
\def\bhat{\hat{b}}
\def\chat{\hat{c}}
\def\that{\hat{t}}

\def\Hcal{\mathcal{H}}

\begin{document}
\title{Low-temperature theory of inversion and quantum oscillations in Kondo insulators}
\author{Arnav Pushkar}
\author{Brijesh Kumar}
\email{bkumar@mail.jnu.ac.in}
\affiliation{School of Physical Sciences, Jawaharlal Nehru University, New Delhi 110067, India.}
\date{\today} 

\begin{abstract}
The half-filled Kondo lattice model is studied at low temperatures on simple cubic lattice using the self-consistent theory developed in Phys. Rev. B {\bf 96}, 075115 (2017). It is found to have three distinct insulating phases in the temperature-hopping plane, namely, the strong coupling Kondo singlet (KS) phase, the inverted Kondo singlet (iKS) phase distinguished from the KS by inversion, and the antiferromagnetic (AFM) phase. The quasiparticle density of states across the inversion transition is noted to exhibit dimensional reduction, which can differentiate between the KS and iKS insulators in experiments. Magnetic quantum oscillations obtained in the iKS and AFM phases are found to show Lifshitz-Kosevich like behaviour with temperature as well as inverse hopping.  
\end{abstract}

%%%%%%%%%%%%%%%%%%% 
\maketitle
\section{Introduction}

Kondo insulators are heavy fermion systems commonly realized in rare-earth compounds~\cite{Varma.RMP.1976, Misra.Book.2008}. They behave as insulators at low temperatures due to the interaction of itinerant electrons with local moments. Recent observations of quantum oscillations in two well-known Kondo insulators, \ce{SmB6}~\cite{li2014two,tan2015unconventional,hartstein2018fermi} and \ce{YbB12}~\cite{Liu_2018,Xiang_2018,Ong_2018}, have led to a surge of interest in the subject. Quantum oscillations of the quantities such as magnetization in response to magnetic field have for long been considered a characteristic of metals~\cite{shoenberg_1984}. These observations challenge that. Different scenarios and theories have been put forward in recent years to understand this intriguing situation of Kondo insulators exhibiting quantum oscillations~\cite{Knolle2015,Baskaran.2015,Zhang2016,Erten2016,Pal2016,Ram2017,Sodemann2018,Ram2019,Peters2019,Tada2020,Devakul2021}.

Of particular interest to us in this paper is a theory of Kondo insulators put forward in Ref.~\cite{Ram2017}. Notably, it finds that the dispersion of the gapped charge quasiparticles undergoes inversion upon decreasing the Kondo coupling, and the quantum oscillations in the insulating bulk appear only after the inversion has occurred. This inversion is a genuine many-body effect resulting from the competition between the Kondo interaction, $J$, and the conduction electron hopping, $t$. For small $t/J$, i.e. strong Kondo couplings, the charge gap comes from the centre of the Brillouin zone. But upon increasing $t/J$ beyond the so-called inversion point, the zone centre becomes a local maxima and the charge gap shifts to a contour around the zone centre.  Across this inversion transition, the ground state remains a Kondo singlet. A further increase in $t/J$ leads to antiferromagnetic ordering. 

This theory elucidates the microscopic basis for the existence of quantum oscillations in the insulating bulk by discovering inversion as a key property of the correlated insulators. It was nicely worked out for two prototypical models of Kondo insulators, 
viz., the half-filled Kondo lattice model~\cite{Ram2017} and the symmetric periodic Anderson model~\cite{Ram2019} in the ground state, i.e. at absolute zero temperature. What happens to the inversion, and the consequent possibility of quantum oscillations, at finite temperatures remains to be understood. It is our goal here to address this question. 

In this paper, we extend the theory enunciated in Ref.~\cite{Ram2017} to finite temperatures, and study the behaviour of inversion and quantum oscillations in the half-filled Kondo lattice model (KLM) on simple cubic lattice. We do the theory in Sec.~\ref{sec:model}, and obtain the phase diagram in the temperature-hopping plane. Notably, we get inversion transition also at finite temperatures, with an `inverted' Kondo singlet phase in an extended region of the phase diagram. We also identify characteristic changes in the quasiparticle density of states and specific heat across inversion transition that can experimentally differentiate between the Kondo insulators distinguished by inversion. In Sec.~\ref{sec:qo}, we study magnetic quantum oscillations at finite temperatures. We get these oscillations in the inverted Kondo singlet as well as antiferromagnetic phases, and find them to follow Lifshitz-Kosevich like behaviour. We conclude this paper with a summary in Sec.~\ref{sec:sum}.

%%%%
\section{\label{sec:model} Half-filled Kondo Lattice Model}
The Kondo lattice model (KLM)
\begin{equation}
\Hcal = -t\sum_{\vec{r},\vec{\delta}}\sum_{s=\uparrow,\downarrow}\chat_{\vec{r},s}^\dag \chat_{\vec{r}+\vec{\delta},s}+\frac{J}{2}\sum_{\vec{r}} \vec{S}_{\vec{r}} \cdot \vec{\tau}_{\vec{r}}
\end{equation}
 of local moments coupled antiferromagnetically ($J>0$) to the electrons of a half-filled conduction band describes Kondo insulators. On bipartite lattices with nearest-neighbour hopping, $t$, exact half-filling is ensured by zero chemical potential. Here, we consider simple cubic lattice formed by $L$ sites with position vectors $\{\vec{r}\}$; $\vec{\delta}$ denotes the nearest-neighbours of every point $\vec{r}$. The local moments are described by the Pauli operators, $\vec{\tau}_{\vec{r}}$'s, while $\chat_{\vec{r},s} (\chat^\dag_{\vec{r},s})$ and $\vec{S}_{\vec{r}}$ denote respectively the annihilation (creation) and spin operators of the conduction electrons. 
 
In Ref.~\cite{Ram2017}, a self-consistent theory of Kondo insulators was formulated in the Kumar representation of electrons~\cite{Kumar2008}: $\chat^\dag_{\vec{r}\uparrow} = \hat{\phi}_{a,\vec{r}} \, \sigma^+_{\vec{r}}$, $\chat^\dag_{\vec{r} \downarrow} = \frac{1}{2}(i\psi_{a,\vec{r}}-\phi_{a,\vec{r}}\,\sigma^z_{\vec{r}})$ on $A$ sublattice and $\chat^\dag_{\vec{r},\uparrow} = i\hat{\psi}_{b,\vec{r}}\,\sigma^+_{\vec{r}}$, $\chat^\dag_{\vec{r},\downarrow} = \frac{1}{2}(\phi_{b,\vec{r}}-i\psi_{b,\vec{r}}\,\sigma^z_{\vec{r}})$ on $B$ sublattice; here $\phi_{a(b),\vec{r}}$, $\psi_{a(b)\vec{r}}$ are Majorana fermions and $\sigma^z_{\vec{r}}$, $\sigma^\pm_{\vec{r}}$ are Pauli operators. Following Ref.~\cite{Ram2017}, the half-filled KLM is written as: $\Hcal\approx \Hcal_c + \Hcal_s +e_0L$, where 
\begin{subequations}
\begin{eqnarray}
\mathcal{H}_c &=&  \frac{J\rho_0}{4}\left(\sum_{\vec{r}\in A}\ahat^\dag_{\vec{r}}\,\ahat^{ }_{\vec{r}}+\sum_{\vec{r}\in B}\bhat^\dag_{\vec{r}}\,\bhat^{ }_{\vec{r}}\right)  \nonumber \\
&& -\frac{it}{2}\sum_{\vec{r}\in A}\sum_{\vec{\delta}}\left(\psi^{ }_{a,\vec{r}}\,\phi_{b,\vec{r}+\vec{\delta}}+\rho_1\psi_{b,\vec{r}+\vec{\delta}} \, \phi^{ }_{a,\vec{r}}\right) \label{eq:Hcr} \\
&=& \frac{J\rho_0L}{8} + \sum_{\vec{k}}\sum_{\nu=\pm} E_{\vec{k},\nu}\left(\eta^\dag_{\vec{k},\nu} \eta^{ }_{\vec{k},\nu} -\frac{1}{2}\right) \label{eq:Hck}
\end{eqnarray}
\end{subequations}
describes the effective `charge' dynamics in terms of the spinless fermions $\ahat_{\vec{r}}$ and $\bhat_{\vec{r}}$ such that $ \phi_{a,\vec{r}} = \ahat^{ }_{\vec{r}} + \ahat^\dag_{\vec{r}}$ and $ \psi_{a,\vec{r}} = i(\ahat^{ }_{\vec{r}} - \ahat^\dag_{\vec{r}})$ on $A$ sublattice, and likewise for $\bhat_{\vec{r}}$ on $B$ sublattice; the parameters $\rho_0$ and $\rho_1$ defined below are to be determined self-consistently. After doing Fourier and Bogoliubov transformations, we derive from Eq.~\eqref{eq:Hcr} the diagonalized form of $\Hcal_c$ in Eq.~\eqref{eq:Hck}, where $\eta_{\vec{k},\nu}$ are the quasiparticle operators corresponding to the $\vec{k}$ points in the half-Brillouin zone, and $E_{\vec{k},\pm} = E_{\vec{k}} \pm +\frac{1}{2}t(1+\rho_1)|\gamma_{\vec{k}}|$ are the quasiparticle  dispersions; here $\gamma_{\vec{k}}=\sum_{\vec{\delta}} e^{i\vec{k}\cdot\vec{\delta}}$ and $E_{\vec{k}}=\sqrt{(J\rho_0/4)^2+(t(1-\rho_1)|\gamma_{\vec{k}}|/2)^2}$. The `spin' physics of KLM is described effectively by the following model.
\begin{subequations}
\begin{align}
\mathcal{H}_s=\frac{J\bar{n}}{4}\sum_{\vec{r}} \vec{\sigma}_{\vec{r}}\cdot\vec{\tau}_{\vec{r}} + \frac{t \zeta}{4}\sum_{\vec{r},\delta}\vec{\sigma}_{\vec{r}} \cdot \vec{\sigma}_{\vec{r}+\vec{\delta}} \label{eq:Hsr}
\end{align}
The four self-consistent parameters of this theory are defined as: $\rho_0=\frac{1}{L}\sum_{\vec{r}} \langle\vec{\sigma}_{\vec{r}}\cdot\vec{\tau}_{\vec{r}}\rangle$, $\rho_1=\frac{1}{zL}\sum_{\vec{r},\vec{\delta}} \langle \vec{\sigma}_{\vec{r}} \cdot \vec{\sigma}_{\vec{r}+\vec{\delta}}\rangle$, $\zeta=\frac{2i}{zL}\sum_{\vec{r}\in A,\vec{\delta}}\langle \phi^{ }_{a,\vec{r}} \psi_{b,\vec{r}+\vec{\delta}}\rangle$, $\bar{n}=\frac{1}{L}\langle\sum_{\vec{r}\in A}\ahat^\dag_{\vec{r}}\ahat^{ }_{\vec{r}}+\sum_{\vec{r}\in B}\bhat^\dag_{\vec{r}}\bhat^{ }_{\vec{r}}\rangle $. The constant term is $e_0=-(J\bar{n}\rho_0 + zt\zeta\rho_1)/4$. We urge the readers to look at Ref.~\cite{Ram2017} for more details.

The $\Hcal_s$ is treated analytically using bond-operator mean-field theory~\cite{sachdev1990bond} in terms of the singlet and triplet eigenstates of the local interaction, $J\vec{\sigma}_{\vec{r}}\cdot\vec{\tau}_{\vec{r}}$. It amounts to writing the Pauli operators approximately as $\sigma_{\vec{r},\alpha}\approx \bar{s}(\that^\dag_{\vec{r},\alpha} +\that^{ }_{\vec{r},\alpha}) \approx -\tau_{\vec{r},\alpha}$ for $\alpha=x,y,z$; here $\bar{s}^2 = (1-\rho_0)/4$ is the weight of Kondo singlet per site, and the boson operators, $\that_{\vec{r},\alpha}$, describe the triplet excitations. By doing Fourier and Bogoliubov transformations, we obtain the following diagonalized $\Hcal_s$ with triplon dispersion, $\epsilon_{\vec{k}}=\sqrt{\lambda(\lambda+t\zeta\bar{s}^2\gamma_{\vec{k}})}$, in the momentum space; here $\lambda$ is the Lagrange multiplier that enforces on average the physical constraint, $\bar{s}^2+\sum_\alpha \that^\dag_{\vec{r},\alpha}\that^{ }_{\vec{r},\alpha}=1$. 
\begin{align}
\Hcal_s \approx L\left[\lambda\bar{s}^2-\frac{5\lambda}{2}+\frac{J\bar{n}\rho_0}{4}\right]+\sum_{\vec{k},\alpha}\epsilon_{\vec{k}}\left(\tilde{t}^\dag_{\vec{k},\alpha}\tilde{t}^{ }_{\vec{k},\alpha}+\frac{1}{2}\right)
\label{eq:Hsk}
\end{align}
\end{subequations}

\subsection{\label{sec:sce-T} Self-consistent equations at finite temperatures}
The parameters, $\rho_0$, $\rho_1$, $\zeta$, $\bar{n}$ of this theory are to be determined self-consistently. In Ref.~\cite{Ram2017}, these calculations were done at temperature $T=0$ only. Here we do so at finite $T$ to understand how with temperature the inversion and quantum oscillations behave in this theory. Equations for $\zeta$ and $\bar{n}$, obtained by thermal averaging the related operators (defined above) with respect to Eq.~\eqref{eq:Hck}, are written below; here $\beta=1/{T}$.
\begin{subequations}
\label{eq:nbar-zeta}
\begin{align}
\bar{n} &= \frac{1}{2}-\frac{J\rho_0}{8L}\sum_{\vec{k},\nu}\frac{\tanh{(\beta E_{\vec{k},\nu}/2)}}{E_{\vec{k}}} \label{eq:nbar} \\
\zeta &= \frac{1}{zL}\sum_{\vec{k},\nu} |\gamma_{\vec{k}}| \left[\frac{t(1-\rho_1)|\gamma_{\vec{k}}|}{2E_{\vec{k}}}-\nu\right] \tanh{(\beta E_{\vec{k},\nu}/2)} \label{eq:zeta}
\end{align}
\end{subequations}
By minimizing with respect to $\bar{s}^2$ and $\lambda$ the free energy, $\mathcal{F}_{s} = -\frac{1}{\beta}\log{\tr{e^{-\beta\Hcal_{s}}}}$, obtained from Eq.~\eqref{eq:Hsk}, we derive the following self-consistent equations.
\begin{subequations}
\begin{align}
\bar{s}^2 &=\frac{5}{2}-\frac{3}{4L}\sum_k\frac{2\lambda+t\zeta\bar{s}^2\gamma_{\vec{k}}}{\epsilon_{\vec{k}}}\coth{(\beta\epsilon_{\vec{k}}/2)}  \label{eq:sbar} \\
\lambda &=J\bar{n}-\frac{3}{4L}\sum_{\vec{k}} \frac{\lambda t \zeta \gamma_{\vec{k}}}{\epsilon_{\vec{k}}} \coth{(\beta\epsilon_{\vec{k}}/2)} \label{eq:lam}
\end{align}
\end{subequations}
From this, we get $\rho_0 = 1-4\bar{s}^2$ and $\rho_1=\frac{4\bar{s}^2(J\bar{n}-\lambda)}{zt\zeta}$. Note that Eqs.~\eqref{eq:nbar} to~\eqref{eq:lam} reduce in the zero temperature limit to the self-consistent equations derived in Ref.~\cite{Ram2017}.

%%%%%%%%%%%%%%%%%%%
\subsection{\label{sec:PD} Phase diagram}
\begin{figure}[htbp]
   \centering
   \includegraphics[width=.8\columnwidth]{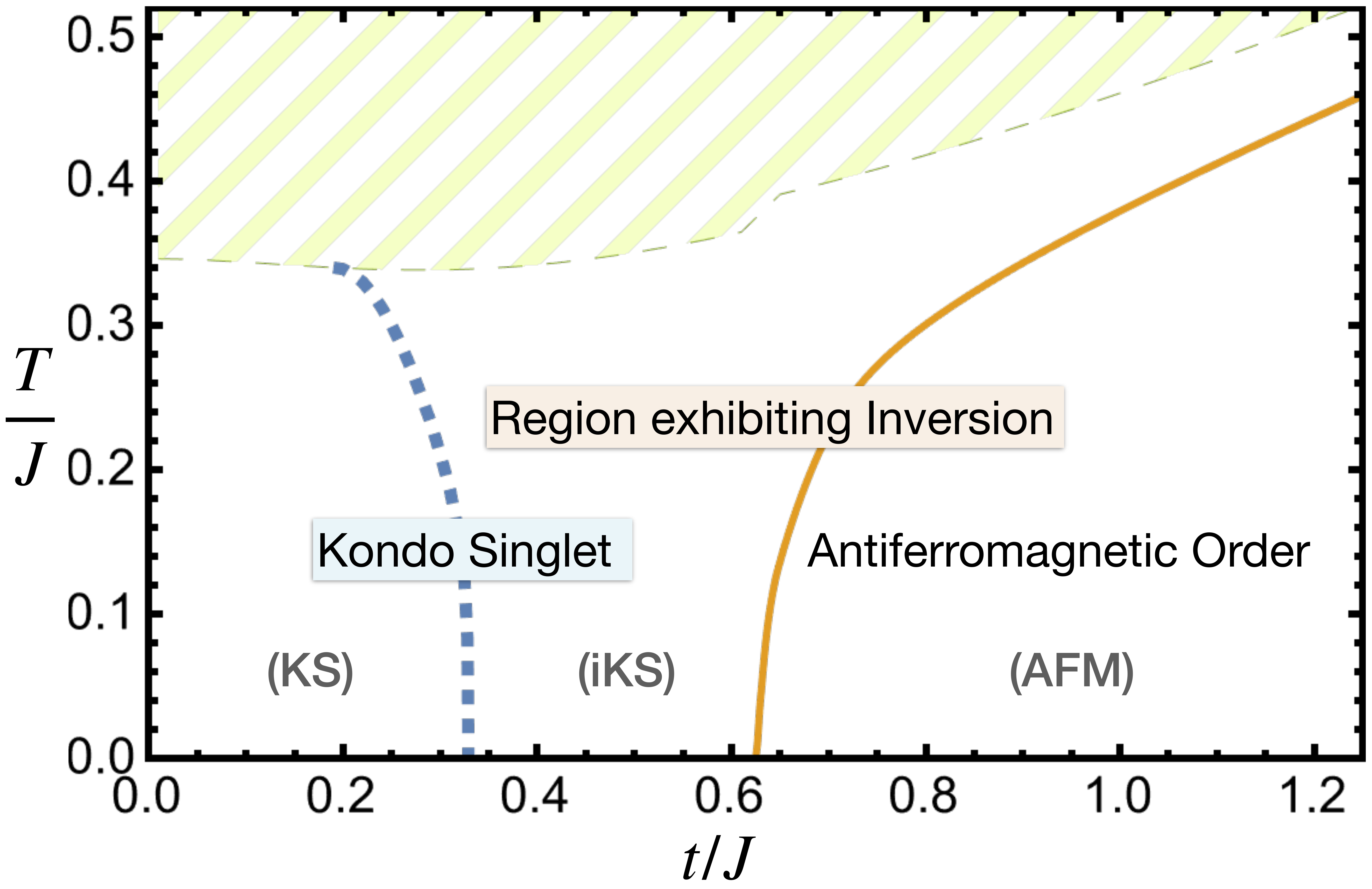}
   \caption{Phase diagram of the half-filled Kondo lattice model in the hopping ($t$)-temperature ($T$) plane in units of Kondo interaction ($J$). It has three low-temperature insulating phases: strong coupling Kondo singlet (KS), inverted Kondo singlet (iKS), and the antiferromagnetic (AFM) phase. Dotted line marks the inversion transition. Continuous line marks the antiferromagnetic transition. Hatched region marks the limitation of the present treatment of $\Hcal_s$ for higher temperatures.}
   \label{fig:PD}
\end{figure}

By solving Eqs.~\eqref{eq:nbar}-\eqref{eq:lam} self-consistently for $\bar{n}$, $\zeta$, $\rho_0$ and $\rho_1$, we obtain the charge and spin dispersions and the respective energy gaps as a function of $t$ and $T$ in units of $J$. At $T=0$, we get the results known from Ref.~\cite{Ram2017}. Namely, upon increasing $t$, the inversion of charge dispersion starts at $t_i=0.33$, and the spin gap closes continuously at $t_c=0.62$ causing a transition from the singlet phase to antiferromagnetic (AFM) phase. For $t<t_i$ the charge gap comes from the zone centre, $\vec{k}=0$, while for $t>t_i$, it comes from the surface, $\gamma_{\vec{k}}=\frac{J\abs{\rho_0}(1-\abs{\rho_1})}{4t(1+\abs{\rho_1})\sqrt{\abs{\rho_1}}}$, that quickly tends to $\gamma_{\vec{k}}=0$ as $t$ increases. By tracing the evolution of the inversion point, $t_i$, and the antiferromagnetic critical point, $t_c$, with temperature, we get a phase diagram in the $t$-$T$ plane, as presented in Fig.~\ref{fig:PD}. 

At low temperatures, we find the half-filled KLM to have three distinct insulating phases separated by the inversion line, $t_i(T)$, and the AFM phase boundary, $t_c(T)$. Across the inversion transition (dotted line in Fig.~\ref{fig:PD}), the spins continue to form  singlet, but the charge dispersion exhibits a change in the form of inversion. Thus, we call the strong coupling singlet phase for $t<t_i$ simply as the Kondo singlet (KS) phase, while that for $t_i<t<t_c$ as the inverted Kondo singlet (iKS) phase; the AFM phase enveloped by the iKS phase lies entirely in the region exhibiting inversion. Notably, the iKS phase is found to grow in extent upon increasing $T$. It implies that most real Kondo insulators (with not-so-strong Kondo couplings) would invariably realise the iKS phase. It has important consequences for quantum oscillations at finite temperatures; as figured out in Ref.~\cite{Ram2017}, inversion is a necessary condition for the quantum oscillation to occur in the Kondo insulating bulk. 

\begin{figure}[t]
\centering
\includegraphics[width=.475\columnwidth]{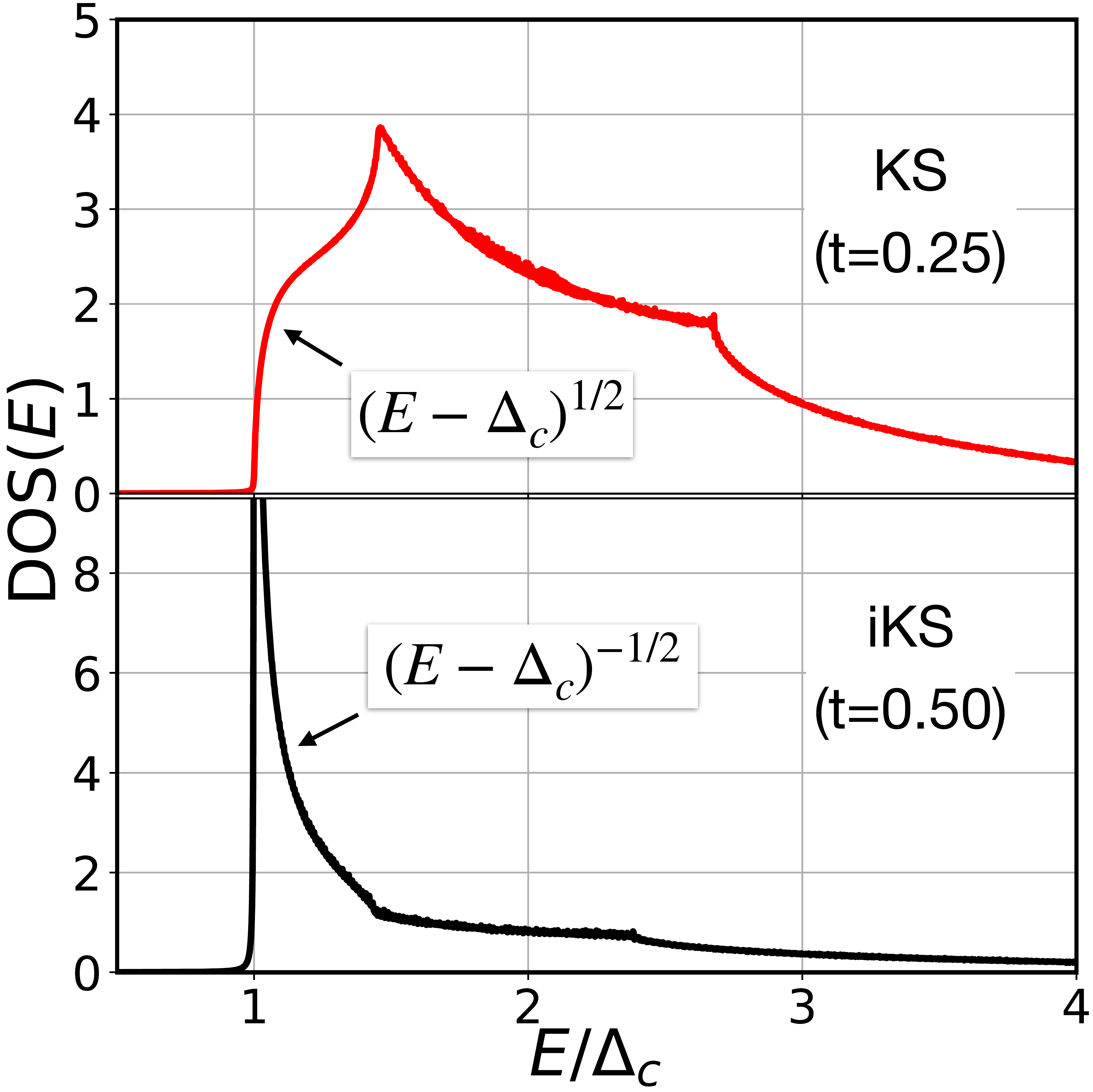}
\hfill \includegraphics[width=.51\columnwidth]{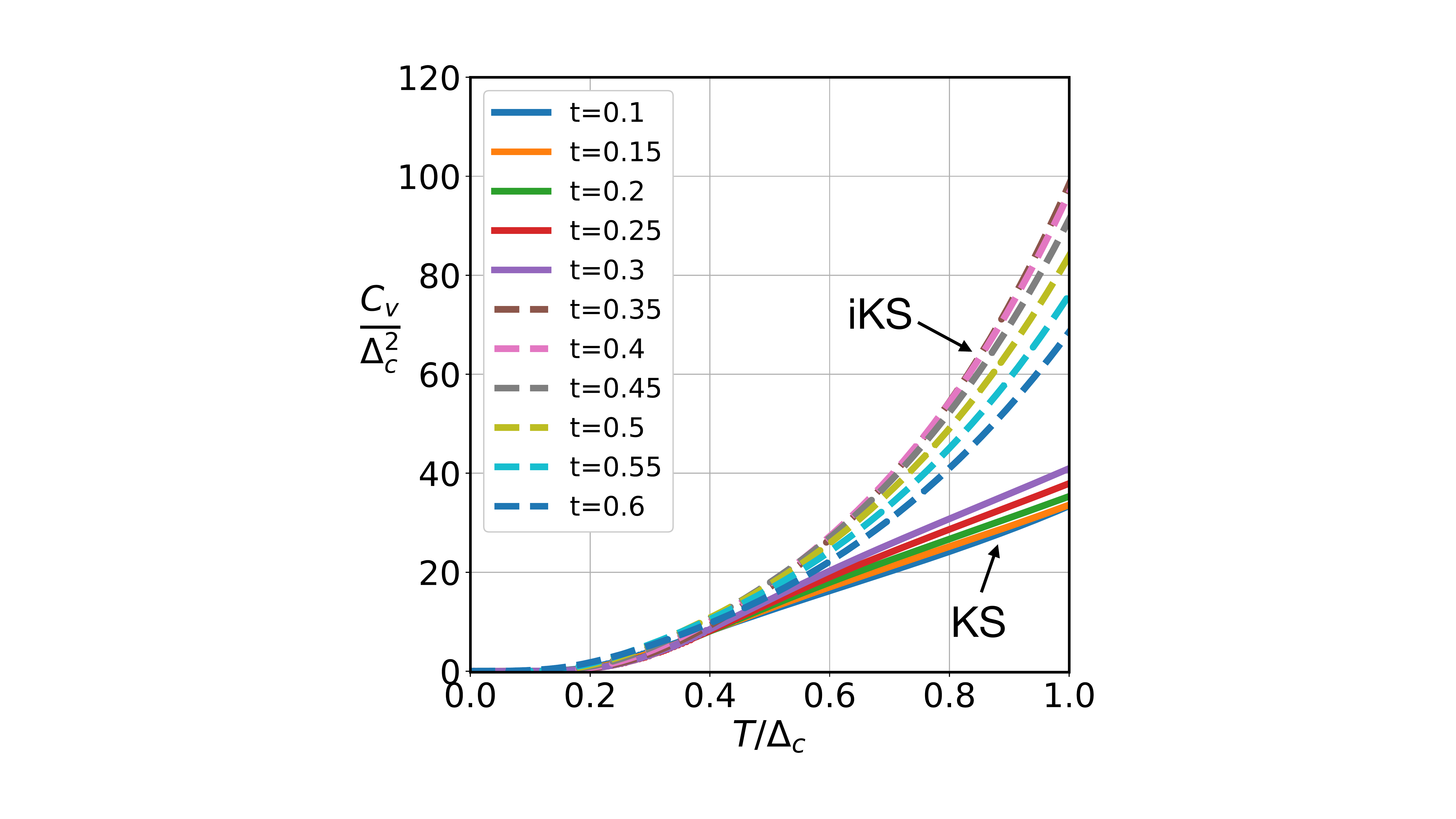}
\caption{({Left}) Quasiparticle density of states, DOS, at $T=0$. (Right) Specific heat, $C_v$. Note the change in behaviour of DOS from $(E-\Delta_c)^{1/2}$ in the KS phase to $(E-\Delta_c)^{-1/2}$ in the iKS phase, near the charge gap $\Delta_c$. It amounts to enhancing $C_v$ in the iKS phase. Also note the bunching of specific heat data for different $t/J$ into two groups corresponding to KS and iKS, when plotted as $C_v/\Delta_c^2$ vs. $T/\Delta_c$.}
\label{fig:dos-cv}
\end{figure}

Across the inversion transition, the quasiparticle density of states (DOS) near the charge gap, $\Delta_c$, exhibits a notable change in the behaviour from $(E-\Delta_c)^{1/2}$ in the KS phase to $ (E-\Delta_c)^{-1/2}$ in the iKS phase; see Fig.~\ref{fig:dos-cv}. It is so caused by the dimensional reduction of dispersion near $\Delta_c$. The low-energy dispersion in the KS phase, $E_{\vec{k},-}-\Delta_c \sim |\vec{k}|^2$, is fully three dimensional around the zone centre, whereas in the iKS phase, $E_{\vec{k},-}-\Delta_c \sim k_\perp^2$ is effectively one dimensional because the energy close to $\Delta_c$ increases only along the normal to the gap surface; $k_\perp$ is the component of $\vec{k}$ $\perp$ to the gap surface. This change in DOS would show up in physical quantities such as the specific heat, $C_v$. In the two types of Kondo insulators distinguished by inversion, the specific heat at low enough temperatures is expected to behave as:  
\begin{align}
C_v & \sim \Delta_c^{2} ~ e^{-\Delta_c/T} \times \left\{ \begin{array}{l} T^{-\frac{1}{2}} ~~~\mbox{(for KS insulators)} \\ T^{-\frac{3}{2}} ~~~\mbox{(for iKS insulators)} \end{array} \right..
\label{eq:Cv}
\end{align}
Since this difference arises due to inversion, it can be used experimentally to differentiate between the two types of Kondo insulators. In both cases, $C_v$ scales as $\Delta_c^2$. Hence, $C_v/\Delta_c^2$ versus $T/\Delta_c$ for different Kondo insulators would form two bunches corresponding to KS and iKS types, as the calculated specific heat in Fig.~\ref{fig:dos-cv} shows. (Notably, the specific heat for iKS insulators has the BCS form~\cite{A-Mermin}; the two behave similarly for the same reason, although they describe different phenomena.)

Upon increasing $T$ further, we hit the boundary of the inaccessible (hatched) region in Fig.~\ref{fig:PD}, whereafter finding self-consistent solutions of Eqs.~\eqref{eq:sbar}-\eqref{eq:lam} becomes difficult. For small $t/J$, it happens where the crossover from the KS to thermal paramagnetic insulator is expected ($T/J \sim 0.375$). For larger $t/J$, it happens where the iKS insulator to metal transition is expected; with extremely slow convergence, our calculations tend to close the charge gap just inside the hatched region, but can not describe the phase beyond it. It leaves room for treating $\Hcal_s$ by other means to approach high temperature phases. The present approach describes the low-temperature insulating phases of the half-filled KLM  in good detail. 

%%%%%
\section{\label{sec:qo} Magnetic Quantum oscillations}
\begin{figure}[htbp]
\centering
\includegraphics[width=.625\columnwidth]{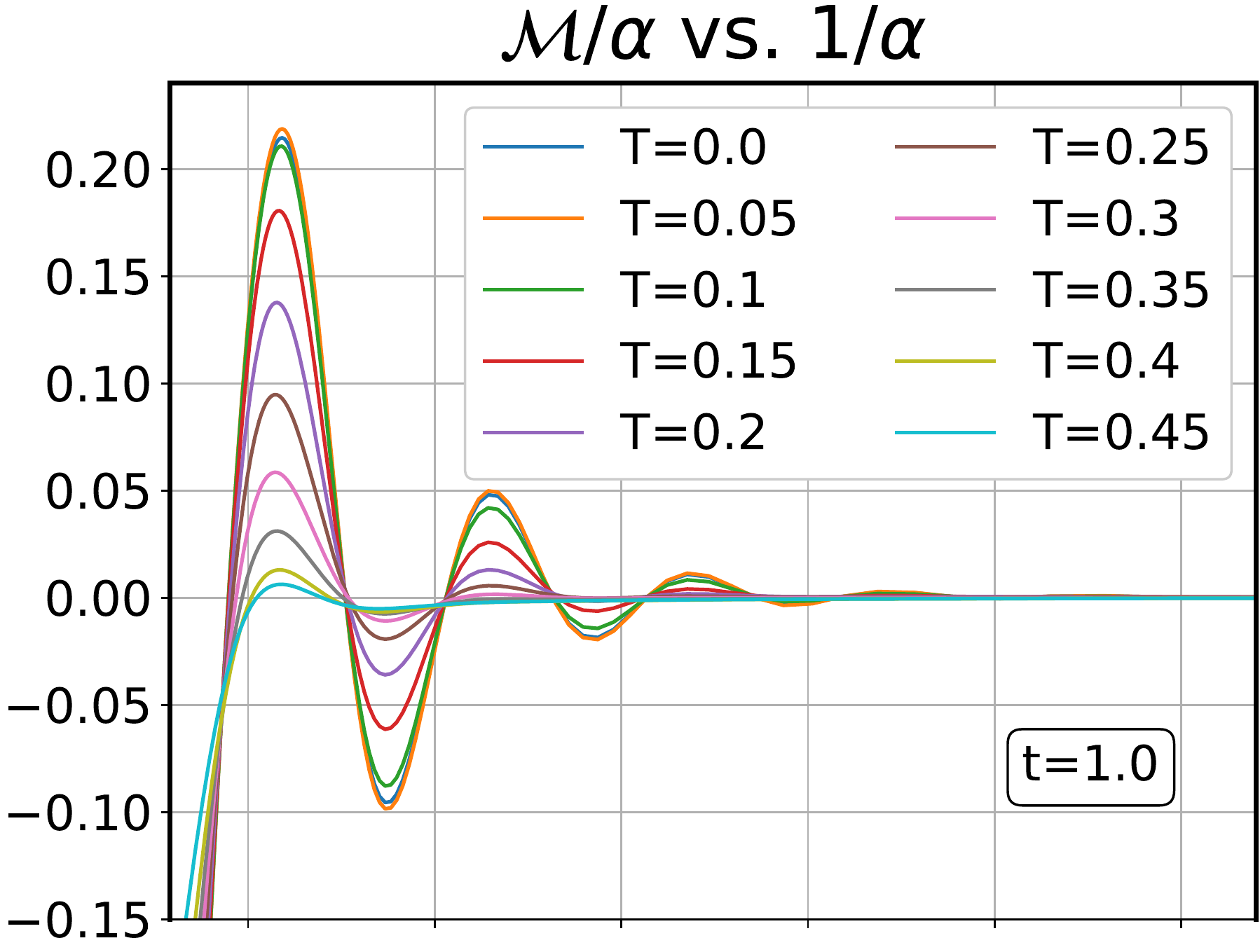} \hfill
\includegraphics[width=.35\columnwidth]{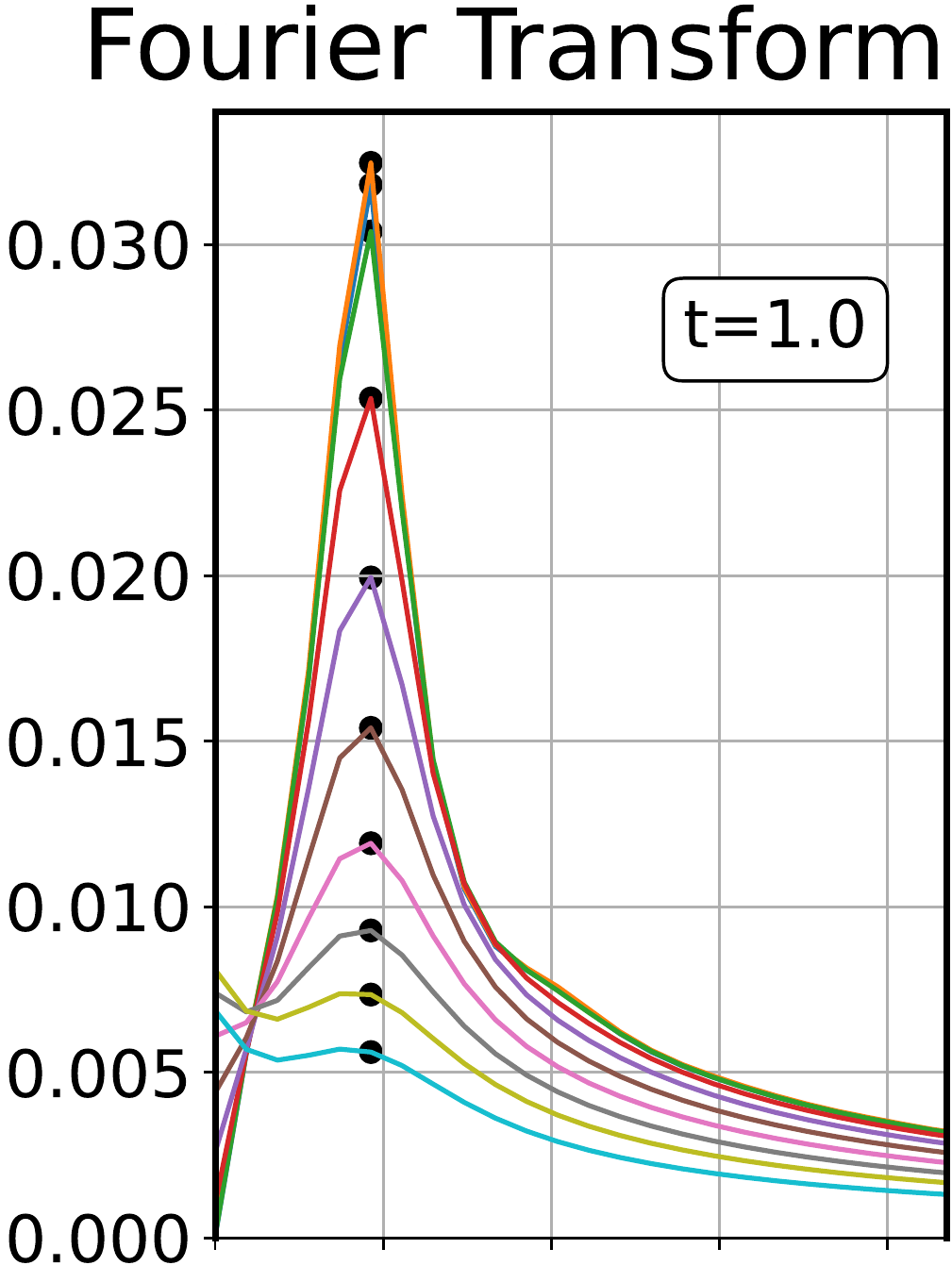}\\
\includegraphics[width=.625\columnwidth]{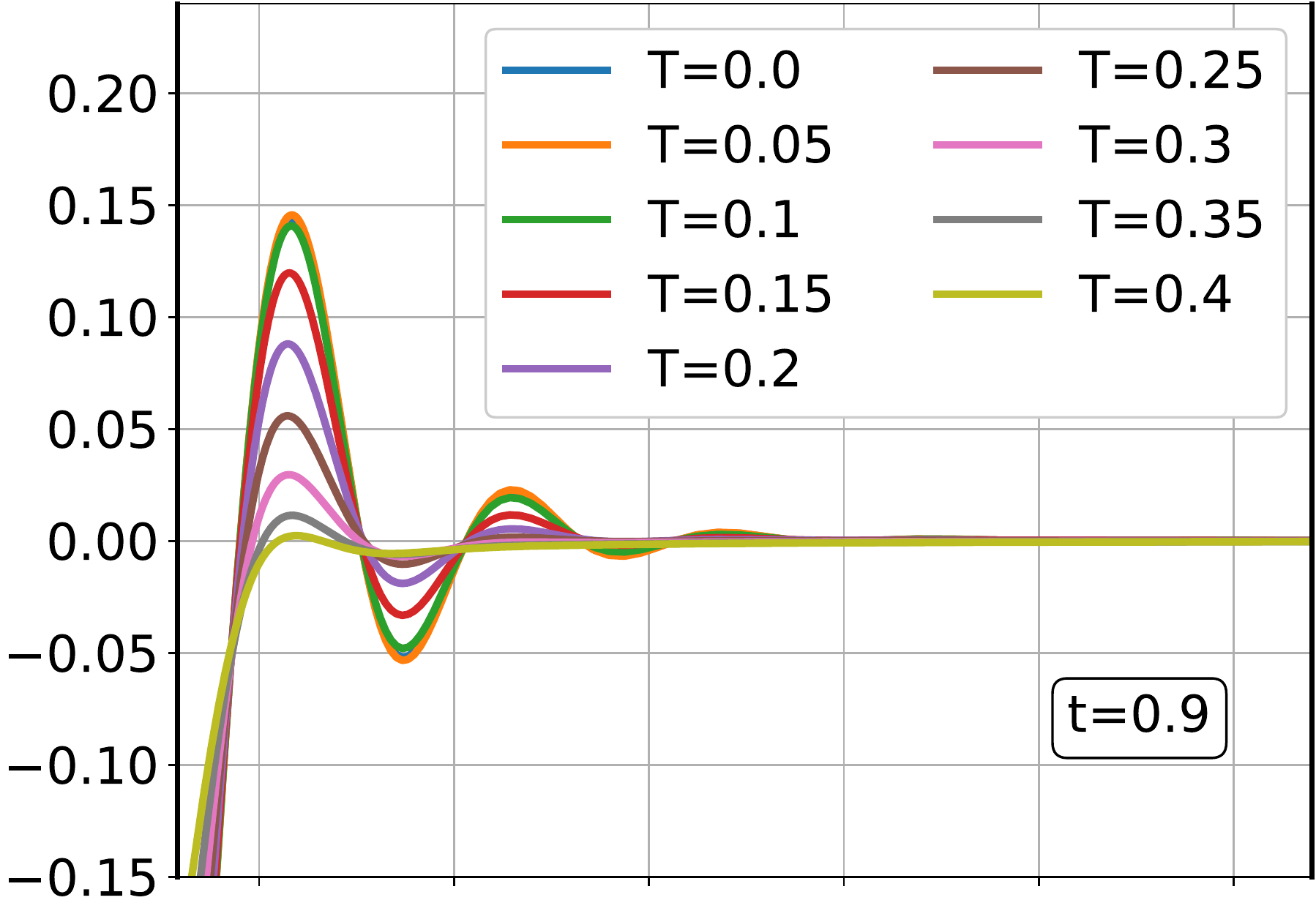} \hfill
\includegraphics[width=.35\columnwidth]{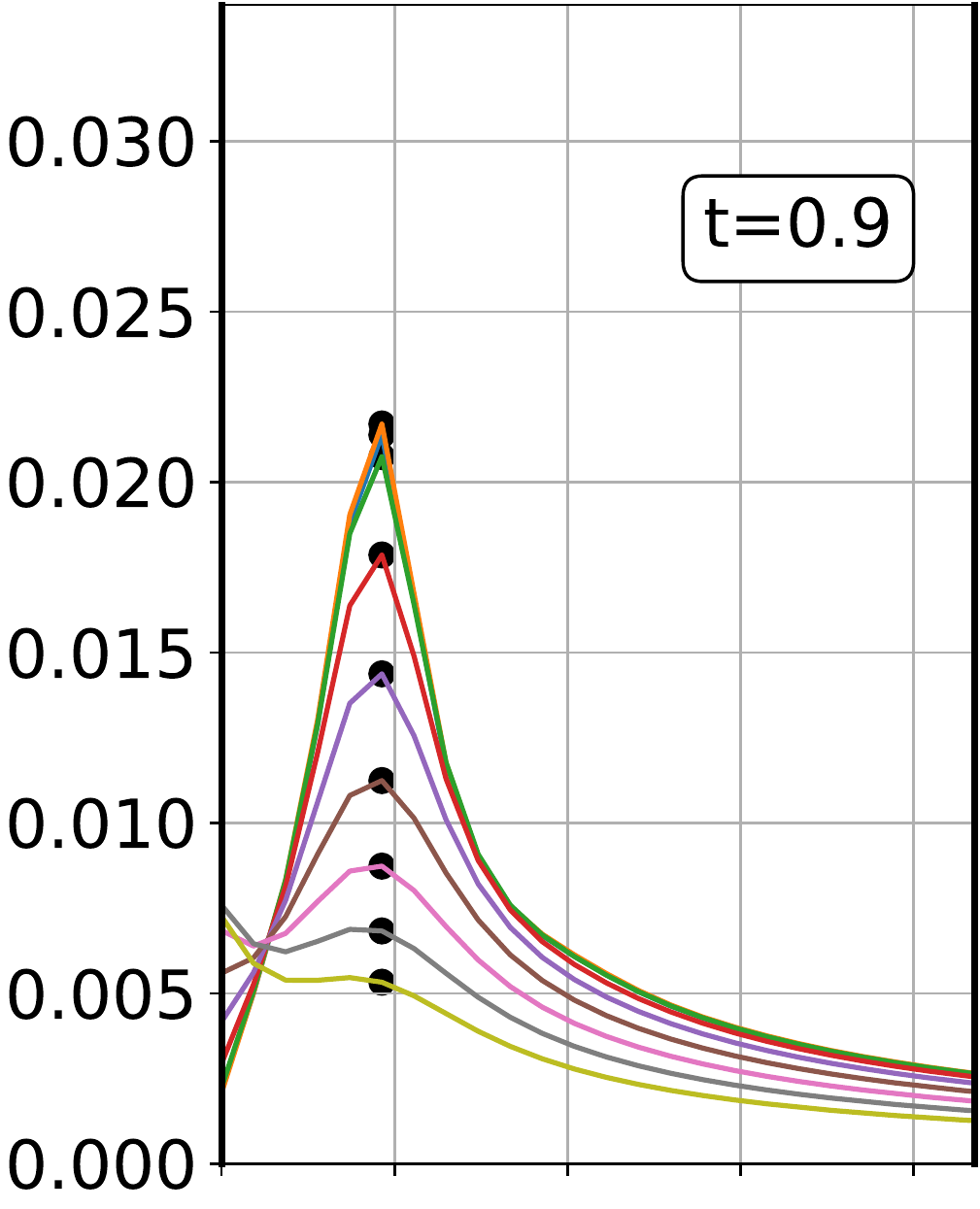}\\
\includegraphics[width=.625\columnwidth]{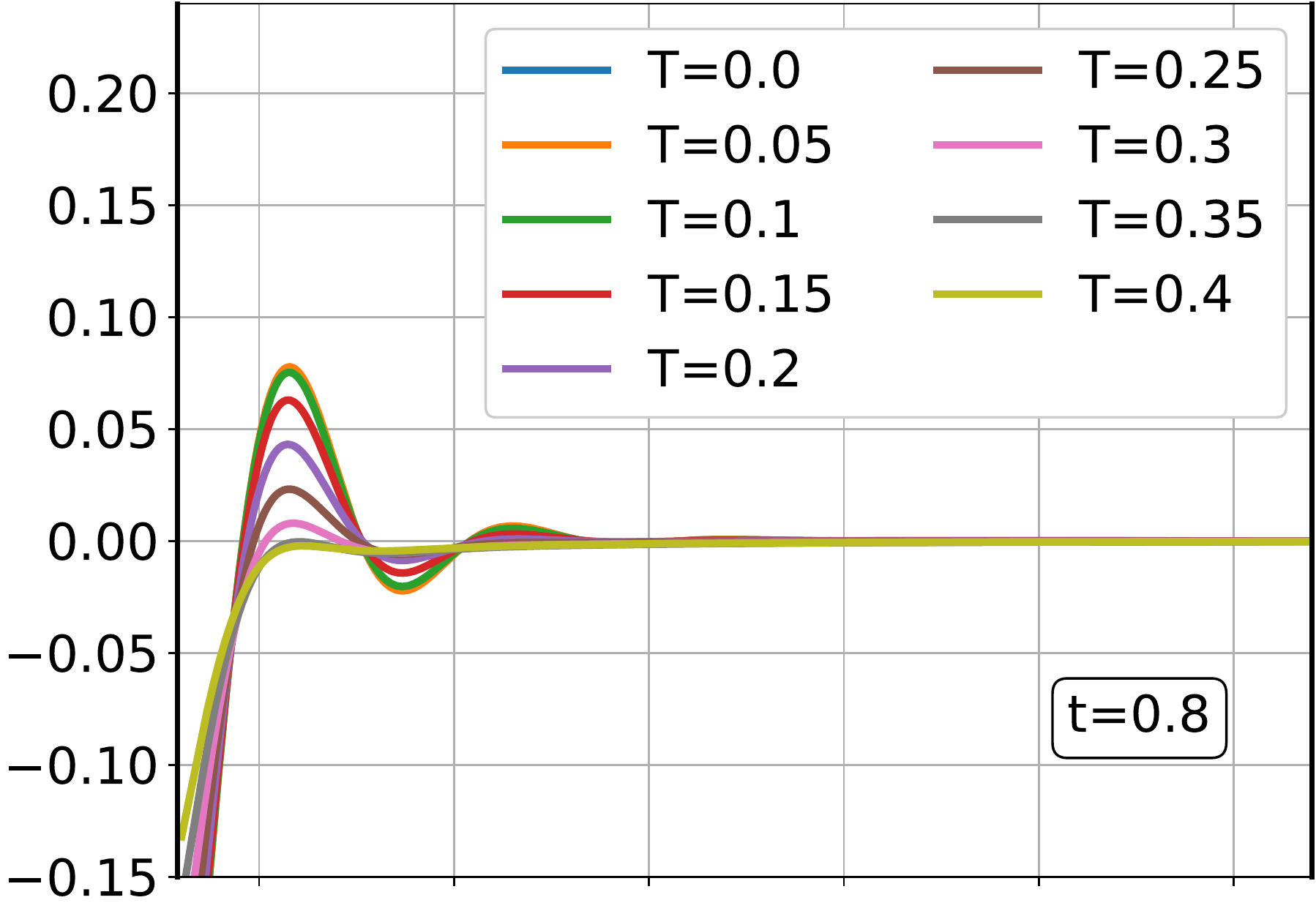} \hfill
\includegraphics[width=.35\columnwidth]{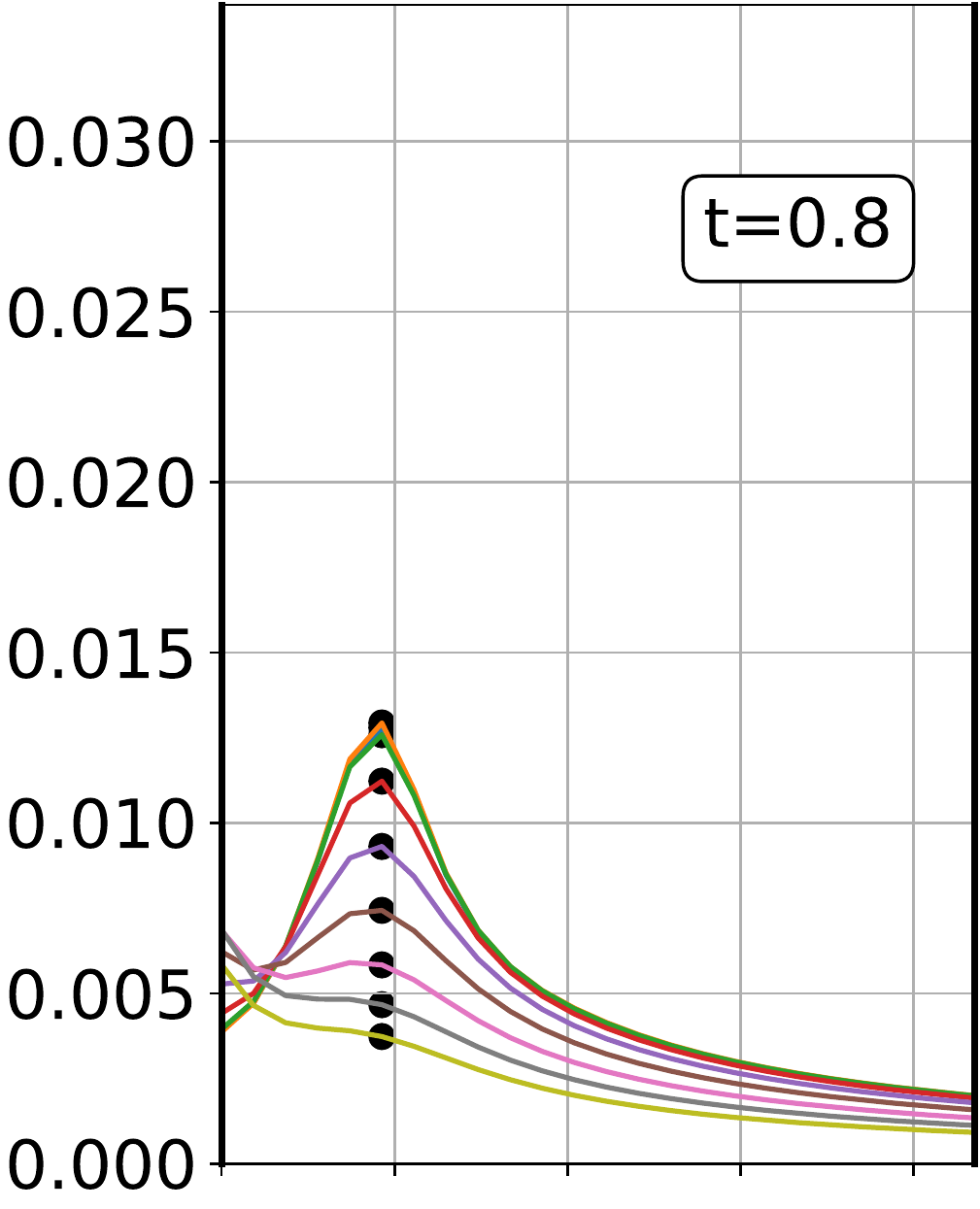}\\
\includegraphics[width=.625\columnwidth]{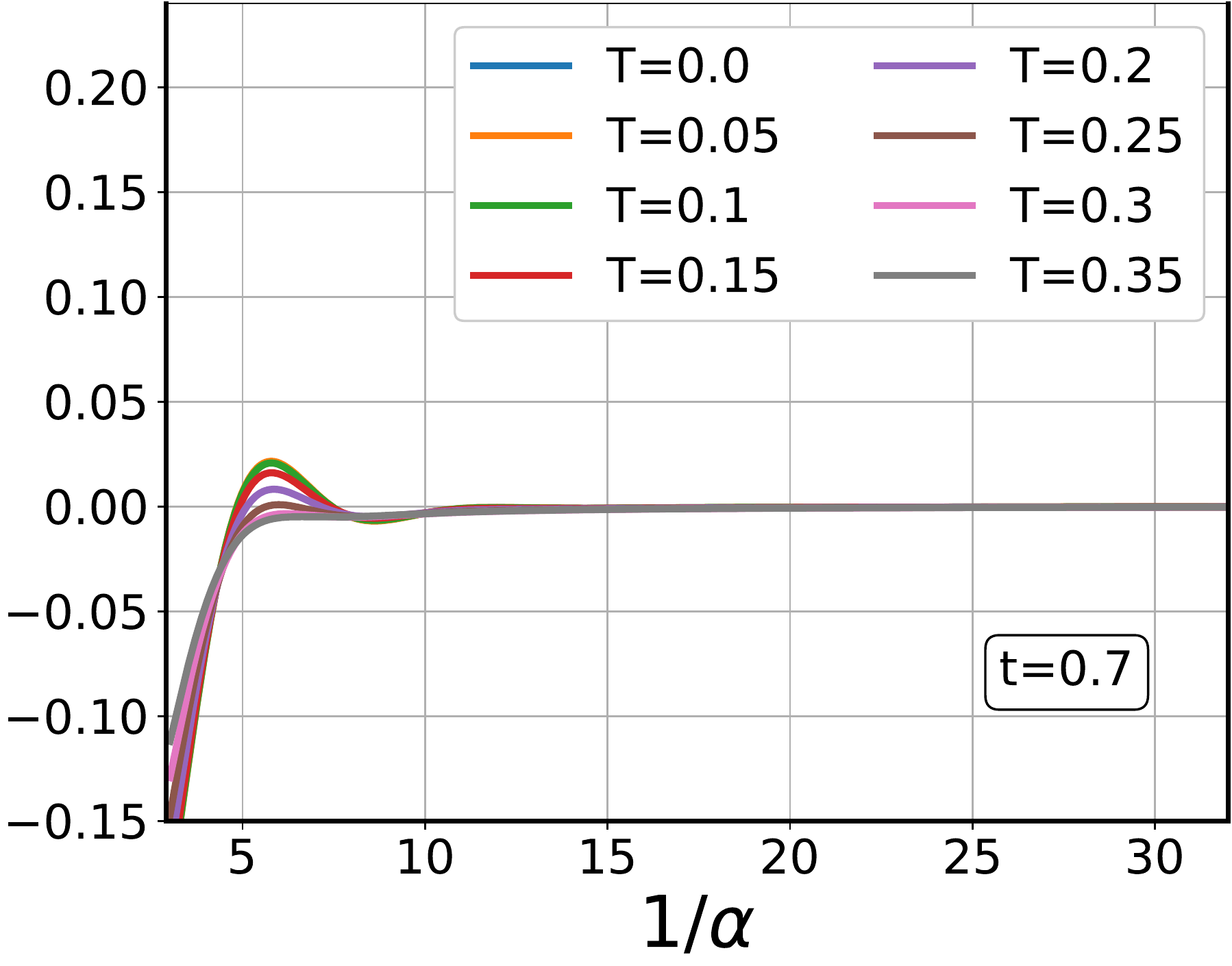} \hfill
\includegraphics[width=.35\columnwidth]{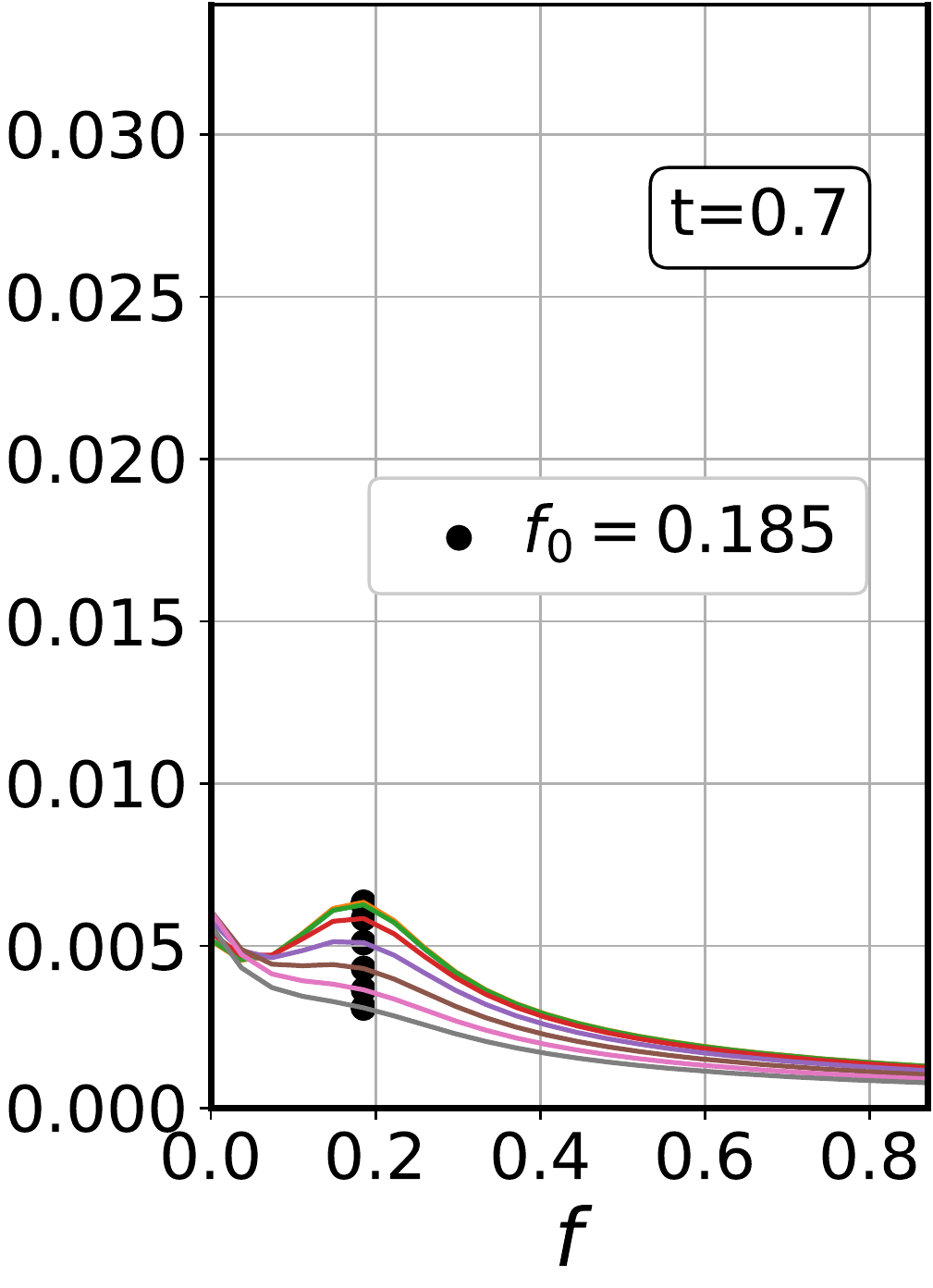}\\
\caption{Magnetic quantum oscillations obtained from our calculations. They oscillate with frequency $f_0=0.185$, and weaken by increasing temperature or decreasing hopping. (The same colour code applies to the data in the plots on the left and the right for the same $t$.) }
\label{fig:qo}
\end{figure}

\begin{figure}[htbp]
\centering
\includegraphics[width=.49\columnwidth]{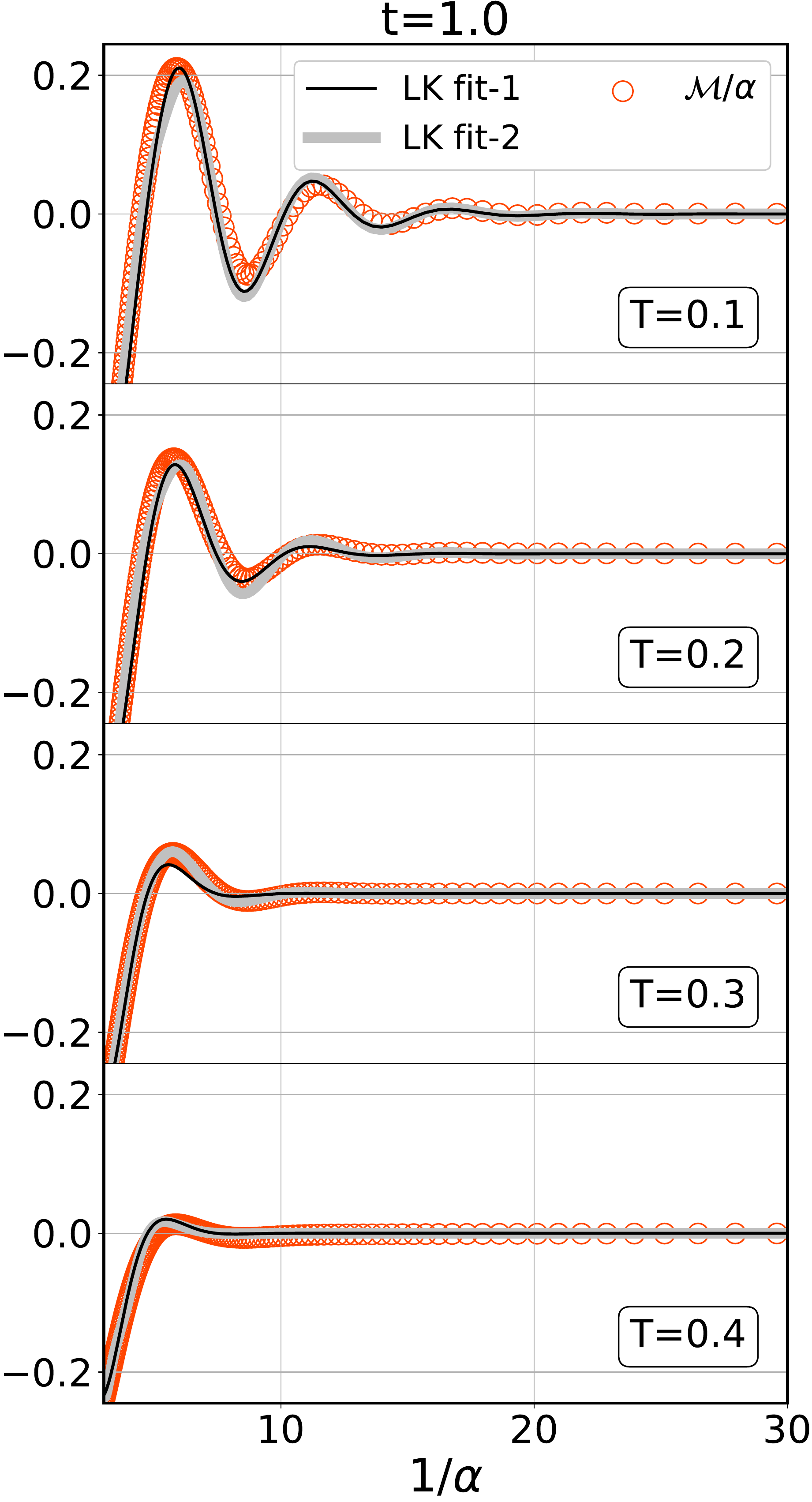}\hfill
\includegraphics[width=.49\columnwidth]{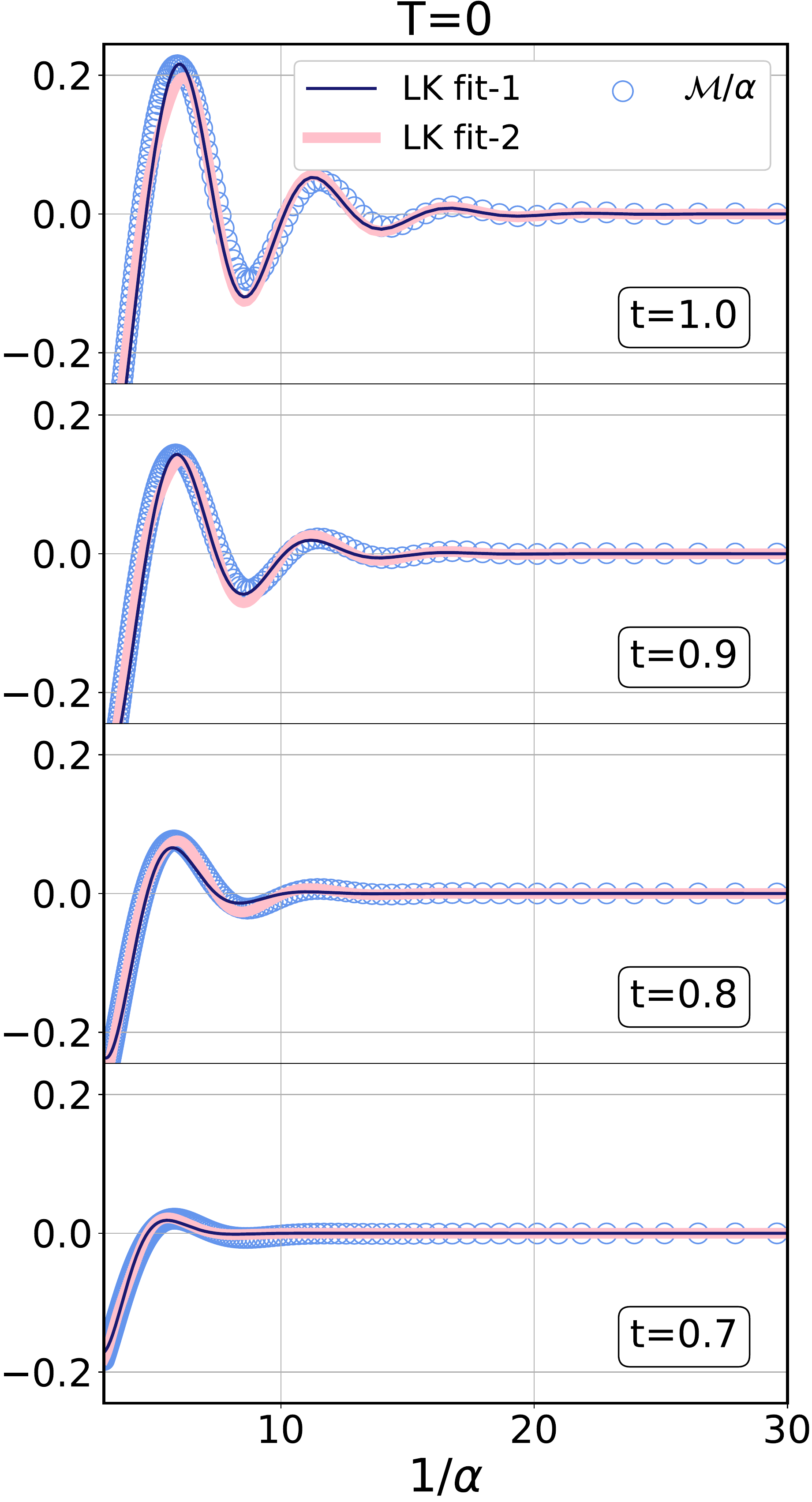}
\caption{Magnetic oscillations fitted with the Lifshitz-Kosevich (LK) form for frequency $f_0=0.185$. A two parameter fit (LK fit-1) with only first two terms of the formula already looks good. LK fit-2 is a five parameter fit involving first four terms.}
\label{fig:LKfit}
\end{figure}

We now study the dynamics of charge quasiparticles in magnetic field, and investigate the behaviour of quantum oscillations at low temperatures. As in Ref.~\cite{Ram2017}, we do it by the following minimal, finite-field extension of the zero-field $\Hcal_c$ of Eq.~\eqref{eq:Hcr}. 
\begin{align}
\mathcal{H}_{c,\alpha} & = \frac{J\rho_0}{4}\left(\sum_{\vec{r}\in A}\ahat^\dag_{\vec{r}}\,\ahat^{ }_{\vec{r}}+\sum_{\vec{r}\in B}\bhat^\dag_{\vec{r}}\,\bhat^{ }_{\vec{r}}\right) -\frac{it}{2}\sum_{\vec{r}\in A}\sum_{\vec{\delta}} \nonumber \\ 
& \cos(2\pi\alpha \, r_y \, \hat{x}\cdot\hat{\delta}) \left[\psi^{ }_{a,\vec{r}}\,\phi_{b,\vec{r}+\vec{\delta}}+\rho_1\psi_{b,\vec{r}+\vec{\delta}} \, \phi^{ }_{a,\vec{r}}\right]
\label{eq:Hcalpha}
\end{align}
Here $\alpha=(Ba^2)/(e/h)$ is the magnetic flux in units of $e/h$, for magnetic field $B$ (along $z$ direction described by the vector potential $\vec{A} = -y B \hat{x}$ along $x$ direction); $a$ is the lattice constant, and $r_y$ is the $y$ component of $\vec{r}$. In numerical calculations, $\alpha$ is taken as $p/q$, with $p=1,2\dots q$ for a prime number $q$; we take, $q=503$. We diagonalize $\Hcal_{c,\alpha}$ numerically for different values of $\alpha$ for several different $T$'s and $t$'s by putting in Eq.~\eqref{eq:Hcalpha} the corresponding zero-field values of $\rho_0$ and $\rho_1$. We then calculate its free energy, $\mathcal{F}_c(\alpha,T,t)$, and obtain from it the magnetization, $\mathcal{M}=-(\partial \mathcal{F}_c/\partial \alpha)/L$. 

In Fig.~\ref{fig:qo}, we present the magnetization data thus calculated. As in the ground state~\cite{Ram2017}, we get oscillations of $\mathcal{M}/\alpha$ with respect to $1/\alpha$ also at finite low temperatures. These oscillations occur only in the inverted region; they grow in strength by increasing $t/J$ and by decreasing $T/J$. For $t/J$ around $0.7$, we begin to see clear magnetic quantum oscillations. The iKS phase at such hopping values occurs invariably at higher temperatures, so the oscillations seen in the iKS phase are weaker compared to those seen in the AFM phase, but they are unmistakably there on both sides of the iKS-AFM phase boundary. Fourier transforming the magnetization data gives the oscillation frequency, $f_0=0.185$; it has been identified to correspond to a contour on the $\gamma_{\vec{k}}=0$ surface in the bulk Brillouin zone~\cite{Ram2017}. We fit the calculated magnetization with the Lifshitz-Kosevich formula, $\mathcal{M}  = \frac{T}{\sqrt{\alpha}} \sum_n c_n (-1)^{n+1} \frac{\sin{[(2\pi n f_0/\alpha) + (\pi/4)]}}{\sqrt{n} \sinh{(n b T/ \alpha)}} $~\cite{LK,shoenberg_1984}; ideally, the $c_n$'s are all equal, say $c$. We find that a fit with only two parameters $b$ and $c$, and the first two terms of the series, already describes these oscillations remarkably well with respect to $T$. See Fig.~\ref{fig:LKfit}. Notably, if we replace $T$ in this formula by $1/t$, it describes very well the oscillations at $T=0$. It suggests that $J^2/t$ acts effectively like temperature. Hence, the Kondo insulating bulk not only exhibits quantum oscillations, but it does so in Lifshitz-Kosevich like manner with $T$ as well as $J^2/t$.

%%%%%%%%
\section{\label{sec:sum} Summary}
We have done a low-temperature theory of inversion and quantum oscillations in the half-filled Kondo lattice model on simple cubic lattice with implications for real Kondo insulators. 
Key takeaways from this study are as follows. The Kondo insulators come in two types, the KS (Kondo singlet) and iKS (inverted Kondo singlet), distinguished by inversion. They can be differentiated by spectral or specific heat measurements, as they have different density of states near the charge gap. The iKS insulators can also realize AFM (antiferromagnetic) order by increasing hopping (say, pressure) or decreasing temperature. Magnetic quantum oscillations occur in the bulk of the inverted Kondo insulators (iKS as well as AFM), and curiously, they follow Lifshitz-Kosevich behaviour with respect to temperature and inverse hopping. 

\acknowledgements{B.K. acknowledges SERB (India) for supporting this research under project grant No. CRG/2019/003251. We also acknowledge the DST-FIST funded HPC cluster at the School of Physical Sciences, JNU for computations.}

\bibliography{KLM_Bibliography.bib}
\end{document}